\documentclass{article}
\usepackage{spconf,amsmath,graphicx,hyperref}
\usepackage{algorithm}
\usepackage{algpseudocode}
\usepackage{multirow}

\usepackage{booktabs}

\title{R2-SVC: TOWARDS REAL-WORLD ROBUST AND EXPRESSIVE ZERO-SHOT SINGING VOICE CONVERSION}
\name{Junjie Zheng$^{\star}$ \qquad Gongyu Chen$^{\star}$ \qquad Chaofan Ding \qquad Zihao Chen$^{\dagger}$ \thanks{$^{\star}$ These authors contributed equally to this work. $^{\dagger}$ Corresponding author.} }
\address{AI Lab, Giant Network }

\begin{document}
%
\maketitle

\begin{abstract}
In real-world singing voice conversion (SVC) applications, environmental noise and the demand for expressive output pose significant challenges. Conventional methods, however, are typically designed without accounting for real deployment scenarios, as both training and inference usually rely on clean data. This mismatch hinders practical use, given the inevitable presence of diverse noise sources and artifacts from music separation. To tackle these issues, we propose R2-SVC, a robust and expressive SVC framework. First, we introduce simulation-based robustness enhancement through random fundamental frequency ($F_0$) perturbations and music separation artifact simulations (e.g., reverberation, echo), substantially improving performance under noisy conditions. Second, we enrich speaker representation using domain-specific singing data: alongside clean vocals, we incorporate DNSMOS-filtered separated vocals and public singing corpora, enabling the model to preserve speaker timbre while capturing singing style nuances. Third, we integrate the Neural Source-Filter (NSF) model to explicitly represent harmonic and noise components, enhancing the naturalness and controllability of converted singing. R2-SVC achieves state-of-the-art results on multiple SVC benchmarks under both clean and noisy conditions. Audio samples are available at: \url{https://c9412600.github.io/svc/index.html}
\end{abstract}

\section{Introduction}
\label{sec:intro}
Singing voice conversion (SVC) aims to transform a singer’s voice into that of another without changing the lyrics or overall musical expression. It has broad applications in dubbing, voice chat, and music production~\cite{10890068, jiang2025ref, wang2024samoye, liu2024zero}. In real-world scenarios, however, robustness is crucial, as practical systems must handle background noise, reverberation, echoes, and artifacts introduced by singing voice separation. Ensuring such robustness is key to producing high-quality vocals while preserving both semantic content and expressive characteristics.

Prior studies explored adversarial learning~\cite{du2022noise, chen2024noise} and data augmentation~\cite{he2024noro, huang2022toward}, but these approaches often generalize poorly to unseen conditions. ASR-based content modeling, from phonetic posteriorgrams (PPG)~\cite{sun2016phonetic} to bottleneck features (BNF)~\cite{wang2021accent}, improves noise robustness but struggles with $F_0$ jitter, reverberation, and echoes, leading to reduced naturalness.

To address this trade-off, we propose \textbf{R2-SVC}, a Robust and Real-World zero-shot SVC system. It integrates three modules—Simulation-based Robustness Enhancement (SRE), Singing-Enhanced Timbre and Style Extractor (SETSE), and Neural Source-Filter (NSF)—to preserve both semantic information and expressiveness in noisy conditions. Our contributions are:
(1) Simulating realistic conditions during training via random $F_0$ perturbation and reverberation/echo artifacts to enhance robustness.
(2) Enriching speaker representation with clean vocals, DNSMOS-filtered~\cite{reddy2022dnsmosp835nonintrusiveperceptual} separated vocals, and public singing corpora, enabling timbre preservation and capturing singer-specific styles.
(3) Incorporating NSF for explicit source–filter decomposition, improving naturalness and controllability in challenging scenarios.
(4) Achieving state-of-the-art results on zero-shot SVC, with evaluations on both clean and noisy real-world test sets confirming the effectiveness and applicability of our approach.

\begin{figure*}
    \centering
    \includegraphics[width=0.8\linewidth]{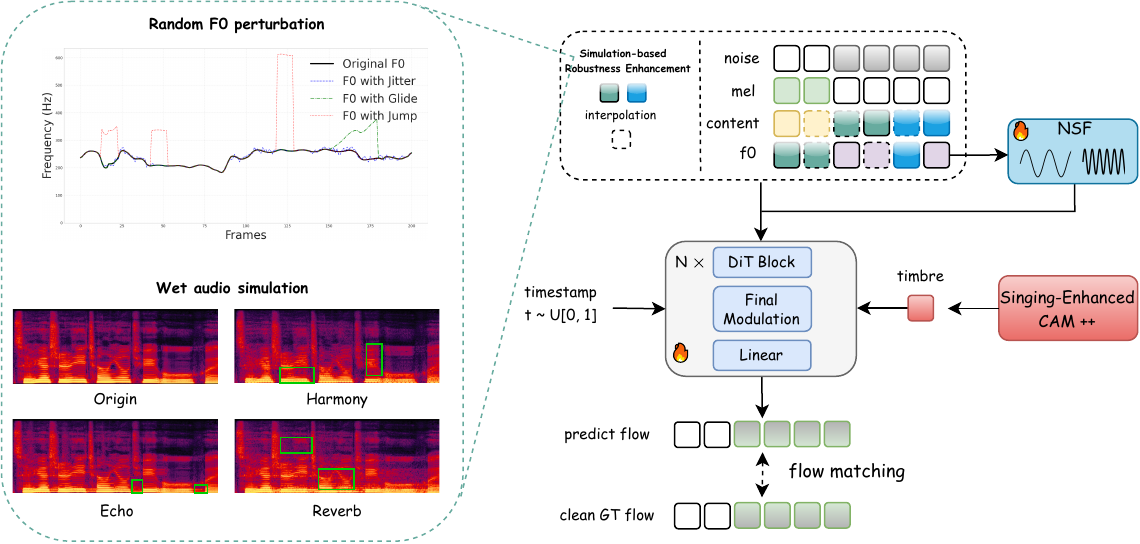}
    \caption{Architecture overview of R2-SVC.}
    \label{fig:overview}
\end{figure*}

\section{Approach}
\label{sec:approach}

\subsection{Overview}
Our approach builds upon a pre-trained Seed-VC checkpoint and is further adapted with open-source singing vocal data to enhance robustness and generalization. Specifically, we introduce three improvements: (i) \emph{simulation-based robustness enhancement}, where random perturbations are applied to the $F_0$ and vocal inputs are augmented with reverberation, harmony and echo artifacts to mimic music separation effects; (ii) \emph{singing-informed timbre and style extractor}, where clean singing vocals, DNSMOS-filtered separated vocals, and public singing corpora are incorporated, enabling the model to not only preserve timbre but also capture speaker-specific singing styles; and (iii) \emph{Neural Source-Filter (NSF) integration}, which provides explicit harmonic and noise representations, improving naturalness and controllability, particularly in singing scenarios.

Formally, the conversion process can be expressed as:
\begin{align}
\hat{y} &= \mathcal{F}_{\theta}\!\left(x^{\mathrm{aug}},\, F_0^{\mathrm{pert}},\, \mathbf{s},\, \mathbf{h}_{\mathrm{nsf}}\right) \\
(x^{\mathrm{aug}},\, F_0^{\mathrm{pert}}) &= \mathcal{R}(x, F_0) \\
\mathbf{s} &= \mathcal{E}(x;\,\mathcal{D}_{\text{sing}}) \\
\mathbf{h}_{\mathrm{nsf}} &= \mathcal{N}(F_0,t) .
\end{align}
Here, $x$ is the source waveform, $x^{\mathrm{aug}}$ and $F_0^{\mathrm{pert}}$ are produced by the robustness module $\mathcal{R}(\cdot)$ via wet-sound simulation and random $F_0$ perturbations; 
$\mathbf{s}$ is a singer-specific timbre/style embedding from the extractor $\mathcal{E}(\cdot)$ fine-tuned on singing data $\mathcal{D}_{\text{sing}}$ (clean and DNSMOS-filtered separated vocals, plus public corpora); and $\mathbf{h}_{\mathrm{nsf}}$ denotes NSF features. 
$\hat{y}$ is the converted singing waveform generated by the conversion model $\mathcal{F}_{\theta}$, which is initialized from the Seed-VC checkpoint and fine-tuned with the augmented singing data.

\subsection{Simulation-based Robustness Enhancement}
In real-world SVC applications, challenges such as inaccurate $F_0$ extraction and residual reverberation or echoes from accompaniment separation are common. To enhance robustness, we propose two strategies: random $F_0$ perturbation and wet sound simulation. The $F_0$ corresponds to the approximate frequency of the quasi-periodic structure in voiced speech, arising from vocal fold oscillations. While incorporating $F_0$ features can improve the pitch accuracy and aesthetic quality of converted audio, extracted $F_0$ contours are often noisy in practice, degrading perceptual quality. A robust model should therefore tolerate such noise and perform implicit corrections to enhance expressiveness.

As shown in Fig.\ref{fig:overview}, we introduce a random $F_0$ perturbation strategy to guide the model’s attention to $F_0$ features. During training, three perturbation types are applied randomly: Jitter, simulating vocal vibrato; Glide, mimicking natural pitch slides; and Jump, emulating abrupt $F_0$ transitions due to prediction errors. This approach reduces the model’s dependency on $F_0$ and encourages reliance on BNF for reconstruction, thereby improving robustness. Vocals separated from accompaniments often contain residual harmonies, reverberation, or echoes. Existing dry audio extraction methods may fail to remove these artifacts without quality loss. To improve model robustness, we simulate such effects using a configurable chain of harmony, echo, and reverb, each triggered with specific probabilities, to augment training data. This teaches the model to produce dry audio from wet inputs. During training, wet sound simulation is applied to all modules except the $F_0$ extractor. Combined with $F_0$ perturbations, this strategy mimics challenging acoustic conditions and strengthens noise tolerance.
\begin{algorithm}[t]
\caption{Training Step with Simulation-based Robustness Enhancement}
\label{alg:train_step_short}
\begin{algorithmic}[1]
\Require Clean waveform $x$; origin pitch contour $F_0$; 
         probs $(0<p_{\text{jit}}+p_{\text{gld}}+p_{\text{jmp}},p_h,p_e,p_r<1)$
\Ensure  Updated model
\State $F_0^{\mathrm{pert}} \gets$ Perturb $F_0$ by \textit{jitter/glide/jump/none} chosen by probs $p_{\text{jit}},p_{\text{gld}},p_{\text{jmp}}$
\State $x^{\mathrm{aug}} \gets$ Apply \textit{harmony}, \textit{echo}, \textit{reverb} independently on $x$ by probs $p_h,p_e,p_r$
\State Feed $(F_0^{\mathrm{pert}},x^{\mathrm{aug}})$ into model
\State Compute loss against clean $x$
\State Update model
\end{algorithmic}
\end{algorithm}

\subsection{Singing-Enhanced Timbre and Style Extractor}
To better capture timbre and stylistic nuances in SVC, we extend the CAM++~\cite{wang2023cam++} framework with a transfer learning strategy using domain-specific singing data. In addition to the original CAM++ datasets, we incorporate public singing voice corpora and high-quality internal vocal data, where tracks are separated and screened with DNSMOS. This augmentation allows the extractor to learn not only the timbre information consistent across speaking and singing, but also singer-specific vocal styles such as vibrato and articulation patterns. By enriching the training data in this way, the extractor adapts more effectively to the singing domain, improving robustness under noisy and reverberant conditions while supporting natural and expressive singing voice conversion.

\subsection{Neural Source-Filter-Based Acoustic Enhancement}
The harmonic-plus-noise Neural Source-Filter (hn-NSF) model \cite{wang2019neuralsourcefilterbasedwaveformmodel} generates waveforms using a source-filter architecture conditioned on acoustic features. Its source module produces a harmonic and noise excitation from $F_0$, which is then spectrally shaped by a filter module. The harmonic component of the excitation, $e(t)$, is generated as follows:
\begin{equation}
\label{eq:source_module}
e(t) = \tanh \left( \mathbf{h}(t) \cdot \mathbf{W}_{\text{merge}}^T + b_{\text{merge}} \right)
\end{equation}
where a vector of sinusoids $\mathbf{h}(t)$, derived from the input frequency $f_0(t)$, undergoes a fixed linear projection and a non-linear $\tanh$ activation. In our framework, this sine-based excitation is further combined with content features to improve naturalness and controllability in SVC.

\begin{figure}[t!]
    \centering
    \includegraphics[width=0.88\linewidth]{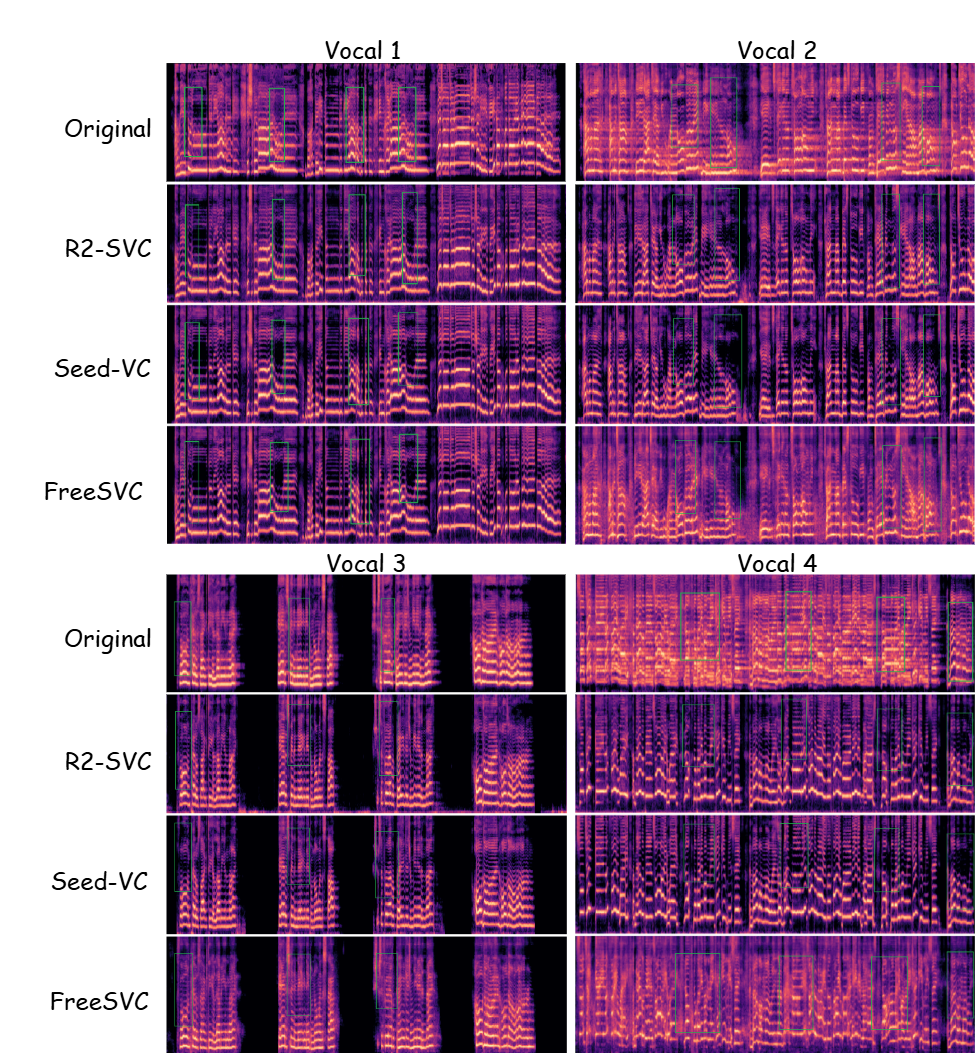}
    \caption{Mel-spectrogram visualizations comparing R2-SVC with Seed-VC and FreeSVC.}
    \label{fig:visualization_of_spec}
\end{figure}

\section{Experiments}
\label{sec:exp}

\begin{table*}[]
\centering
\renewcommand{\arraystretch}{1}
\resizebox{\textwidth}{!}{%
\begin{tabular}{ccccccccccccccccc}
\toprule
 & \multicolumn{8}{c}{SVC-Normal} & \multicolumn{8}{c}{SVC-Hard} \\ \cmidrule(lr){2-9} \cmidrule(lr){10-17}
 & \multirow{2}{*}{SPK-SIM $\uparrow$} & \multirow{2}{*}{CER $\downarrow$} & \multicolumn{4}{c}{Aesthetics} & \multirow{2}{*}{NMOS $\uparrow$} & \multirow{2}{*}{SMOS $\uparrow$} & \multirow{2}{*}{SPK-SIM $\uparrow$} & \multirow{2}{*}{CER $\downarrow$} & \multicolumn{4}{c}{Aesthetics} & \multirow{2}{*}{NMOS $\uparrow$} & \multirow{2}{*}{SMOS $\uparrow$} \\ \cmidrule{4-7} \cmidrule{12-15}
 &  &  & CE $\uparrow$ & CU $\uparrow$ & PC $\uparrow$ & PQ $\uparrow$ &  &  &  &  & CE $\uparrow$ & CU $\uparrow$ & PC $\uparrow$ & PQ $\uparrow$ &  &  \\ \midrule
Original Vocal & - & - & 5.73 & 6.31 & 1.80 & 7.21 & - & - & - & - & 6.31 & 6.81 & 3.34 & 7.44 & - & - \\ \midrule
Seed-VC\cite{liu2024zero} & 71.33\% & \textbf{12.18\%} & 5.43 & 5.97 & 1.66 & 7.08 & 2.75±0.06 & 1.90±0.22 & 72.21\% & \textbf{25.66\%} & 5.57 & 6.04 & 1.83 & 7.20 & 2.79±0.17 & 1.21±0.02 \\
FreeSVC\cite{10890068} & 64.09\% & 14.30\% & 5.15 & 5.74 & \textbf{1.70} & 6.90 & 1.88±0.17 & 1.33±0.01 & 69.50\% & 39.09\% & 4.43 & 5.35 & \textbf{2.05} & 6.91 & 1.44±0.09 & 1.14±0.01 \\
R2-SVC & \textbf{75.17\%} & 13.41\% & \textbf{5.49} & \textbf{5.99} & 1.69 & \textbf{7.08} & \textbf{2.75±0.09} & \textbf{2.06±0.09} & \textbf{76.26\%} & 25.77\% & \textbf{5.63} & \textbf{6.08} & 1.80 & \textbf{7.24} & \textbf{2.90±0.11} & \textbf{2.64±0.10} \\ \midrule
w/o SETSE & 78.95\% & 12.84\% & 5.42 & 5.93 & 1.66 & 7.00 & 2.66±0.11 & 2.02±0.08 & 80.70\% & 28.02\% & 5.58 & 6.03 & 1.75 & 7.24 & 2.75±0.03 & 2.19±0.07 \\
w/o SRE & 74.85\% & 13.18\% & 5.50 & 6.01 & 1.65 & 7.13 & 2.67±0.11 & 2.00±0.13 & 76.17\% & 27.13\% & 5.64 & 6.07 & 1.74 & 7.30 & 2.23±0.21 & 1.74±0.11 \\
w/o NSF & 74.54\% & 12.73\% & 5.54 & 6.05 & 1.66 & 7.17 & 2.55±0.38 & 1.80±0.23 & 75.56\% & 26.17\% & 5.61 & 6.06 & 1.75 & 7.28 & 2.43±0.07 & 1.65±0.08 \\ \bottomrule
\end{tabular}
}
\caption{Objective and subjective evaluation results on the \textbf{SVC-Normal} and \textbf{SVC-Hard} test sets. 
Ablation models include \textbf{SETSE} (Singing-Enhanced Timbre and Style Extractor), \textbf{SRE} (Simulation-based Robustness Enhancement), and \textbf{NSF} (Neural Source-Filter). 
Bold numbers indicate the best results within each column.}
\label{tab:objective_ablation}
\end{table*}

\subsection{Dataset}
We collected six open singing datasets: GTSinger~\cite{zhang2024gtsinger}, M4Singer~\cite{zhang2022m4singer}, OpenCpop~\cite{wang2022opencpop}, OpenSinger~\cite{huang2021multi}, PopBuTFy~\cite{liu2022learning}, and PopCS~\cite{liu2022diffsinger}, totaling around 230 hours. We constructed two test sets: a normal set and a hard set. The normal set includes 100 audio samples randomly selected from our internal test dataset and open singing data. 
The hard test set is constructed to reflect real industrial production scenarios, where even state-of-the-art separation models such as Mel-band Roformer~\cite{wang2023mel} encounter various challenges, including reverberation, echoes, background accompaniment, harmony vocals, etc. 
For target speakers in this task, we randomly selected 10 speakers from our internal test dataset. It is important to note that all test sets and target speakers were unseen during training.

\subsection{Implementation Details}
R2-SVC is built upon an open-source SVC implementation, Seed-VC \cite{liu2024zero}, which utilizes a diffusion transformer (DiT)-based flow matching architecture. The model consists of 17 layers, 12 attention heads, an embedding dimension of 768, and an FFN dimension of 3072. The audio is processed at a sampling rate of 44.1 kHz, using a 2048-point FFT window, a hop size of 512, and 128 mel-frequency bins. All models were initialized from pretrained checkpoints and further trained with an effective batch size of 24 for 50,000 steps. We use the AdamW optimizer with a peak learning rate of $1\times10^{-4}$, which exponentially decays to a minimum of $1\times10^{-5}$. Finally, we employ the pretrained BigVGAN\footnote{https://huggingface.co/nvidia/bigvgan\_v2\_44khz\_128band\_512x} \cite{lee2023bigvgan} model as the vocoder.

For random $F_0$ perturbations, $p_{\text{jit}},p_{\text{gld}},p_{\text{jmp}}$ are set to 0.15, 0.15, and 0.2, respectively. Each audio sample contains between 2 and 4 perturbed segments. In wet audio simulation, the mixing ratios for harmony, echo, and reverb are set to 0.4, 0.35, and 0.5, with setting $p_h,p_e,p_r$ to 0.3, 0.4, and 0.4, respectively. The activation of these three effects is independent. It should also be noted that the model is trained using the perturbed data as input, while the training target remains the clean (unperturbed) data.

\subsection{Baselines}
We compare R2-SVC with two other systems: Seed-VC\footnote{https://github.com/Plachtaa/seed-vc} and FreeSVC\footnote{https://github.com/freds0/free-svc}. As one of the state-of-the-art open-source voice conversion systems, we use the official 200M-parameter Seed-VC checkpoint\footnote{https://huggingface.co/Plachta/Seed-VC/tree/main}, which has been specifically trained for singing voice conversion.

\subsection{Evaluation Metrics}
For objective evaluation, we assess three aspects: speaker similarity, intelligibility, and aesthetic quality. Speaker similarity is quantified using speaker embedding cosine similarity (SPK-SIM), while intelligibility is measured by character error rate (CER). Specifically, we employ Resemblyzer\footnote{https://github.com/resemble-ai/Resemblyzer} to compute SPK-SIM and FireRedASR\footnote{https://github.com/FireRedTeam/FireRedASR}, which was trained on data that includes singing, to obtain CER. Additionally, we introduce an aesthetic score evaluated with the Unified Automatic Quality Assessment model from Meta’s Audiobox Aesthetics\footnote{https://github.com/facebookresearch/audiobox-aesthetics}. For subjective evaluation, we adopt Mean Opinion Score (MOS) tests to measure speaker similarity (SMOS) and speech naturalness (NMOS).

\section{Results}
\subsection{Objective Evaluation}
Table~\ref{tab:objective_ablation} presents results on the SVC-Normal and SVC-Hard test sets. Compared with Seed-VC and FreeSVC, our proposed R2-SVC consistently outperforms or matches baselines. 
It is noteworthy that the results on the hard set show notably worse CER scores compared to the normal set, which is mainly due to the presence of samples with ambiguous pronunciation and more complex singing techniques in the hard set. 
On SVC-Normal, R2-SVC achieves the highest SPK-SIM (75.17\%), competitive CER (13.41\%), and the best CE, CU, and PQ. On SVC-Hard, it further improves SPK-SIM to 76.26\% and attains the best CE (5.63), CU (6.08), and PQ (7.24), showing strong robustness under noisy conditions. While its CER (25.77\%) is slightly above Seed-VC (25.66\%), it is far below FreeSVC (39.09\%), demonstrating a better trade-off between intelligibility and expressiveness.

\subsection{Subjective Evaluation}

In terms of speaker similarity, our model demonstrates performance comparable to that of Seed-VC in zero-shot scenarios, while significantly outperforming Free-SVC. Regarding speech naturalness, our model achieves a notable improvement over Seed-VC. By leveraging a singing-augmented timbre and style extractor along with data augmentation strategies, the converted speech effectively addresses issues such as loss of detail and pitch distortion, leading to a substantial enhancement in naturalness. As illustrated in Figure \ref{fig:visualization_of_spec}, the visualizations of the mel-spectrograms reveal that our model excels in reconstructing fine details compared to the baseline models.

\subsection{Ablation Study}
In addition to baseline comparisons with Seed-SVC~\cite{liu2024zero} and Free-SVC~\cite{10890068}, R2-SVC achieves consistently higher NMOS and SMOS, especially on the challenging ``Hard'' sets, confirming its advantage in naturalness and timbre preservation. Ablation studies further show that while SRE may slightly affect clean data, all three modules—SRE, SETSE, and NSF—contribute significantly to robustness, timbre consistency, and speaker similarity, with R2-SVC yielding the best overall performance under noisy real-world conditions.

\section{Conclusion}
\label{sec:conclu}
We presented \textbf{R2-SVC}, a Real-World Robust and expressive zero-shot singing voice conversion framework that combines simulation-based robustness enhancement, a singing-informed timbre and style extractor, and NSF-based acoustic modeling. Experiments on both clean and noisy test sets demonstrate that R2-SVC achieves state-of-the-art performance while preserving naturalness and expressiveness. In future work, we will explore reinforcement learning for adaptive optimization, improved timbre similarity modeling, and faster inference for real-time applications.

\bibliographystyle{temp/IEEEbib}
\bibliography{strings,refs}

\begin{thebibliography}{10}

\bibitem{10890068}
Alef~Iury Ferreira, Lucas~Rafael Gris, Augusto Da~Rosa, et~al.,
\newblock ``Freesvc: Towards zero-shot multilingual singing voice conversion,''
\newblock in {\em ICASSP}, 2025, pp. 1--5.

\bibitem{jiang2025ref}
Yuepeng Jiang, Ziqian Ning, Shuai Wang, et~al.,
\newblock ``Ref-vc: Robust, expressive and fast zero-shot voice conversion with diffusion transformers,''
\newblock {\em arXiv preprint arXiv:2508.04996}, 2025.

\bibitem{wang2024samoye}
Zihao Wang, Le~Ma, Yongsheng Feng, Yuhang Jin, Kejun Zhang, et~al.,
\newblock ``Samoye: Zero-shot singing voice conversion model based on feature disentanglement and enhancement,''
\newblock 2024.

\bibitem{liu2024zero}
Songting Liu,
\newblock ``Zero-shot voice conversion with diffusion transformers,''
\newblock {\em arXiv preprint arXiv:2411.09943}, 2024.

\bibitem{du2022noise}
Hongqiang Du, Lei Xie, and Haizhou Li,
\newblock ``Noise-robust voice conversion with domain adversarial training,''
\newblock {\em Neural Networks}, vol. 148, pp. 74--84, 2022.

\bibitem{chen2024noise}
Lele Chen, Xiongwei Zhang, Yihao Li, and Meng Sun,
\newblock ``Noise-robust voice conversion using adversarial training with multi-feature decoupling,''
\newblock {\em Engineering Applications of Artificial Intelligence}, vol. 131, pp. 107807, 2024.

\bibitem{he2024noro}
Haorui He, Yuchen Song, Yuancheng Wang, Haoyang Li, Xueyao Zhang, Li~Wang, Gongping Huang, Eng~Siong Chng, and Zhizheng Wu,
\newblock ``Noro: A noise-robust one-shot voice conversion system with hidden speaker representation capabilities,''
\newblock {\em arXiv preprint arXiv:2411.19770}, 2024.

\bibitem{huang2022toward}
Chien-Yu Huang, Kai-Wei Chang, and Hung-Yi Lee,
\newblock ``Toward degradation-robust voice conversion,''
\newblock in {\em ICASSP}. IEEE, 2022, pp. 6777--6781.

\bibitem{sun2016phonetic}
Lifa Sun, Kun Li, Hao Wang, et~al.,
\newblock ``Phonetic posteriorgrams for many-to-one voice conversion without parallel data training,''
\newblock in {\em ICME}. IEEE, 2016, pp. 1--6.

\bibitem{wang2021accent}
Zhichao Wang, Wenshuo Ge, Xiong Wang, et~al.,
\newblock ``Accent and speaker disentanglement in many-to-many voice conversion,''
\newblock in {\em ISCSLP}. IEEE, 2021, pp. 1--5.

\bibitem{reddy2022dnsmosp835nonintrusiveperceptual}
Chandan K~A Reddy, Vishak Gopal, and Ross Cutler,
\newblock ``Dnsmos p.835: A non-intrusive perceptual objective speech quality metric to evaluate noise suppressors,'' 2022.

\bibitem{wang2023cam++}
Hui Wang, Siqi Zheng, Yafeng Chen, Luyao Cheng, and Qian Chen,
\newblock ``Cam++: A fast and efficient network for speaker verification using context-aware masking,''
\newblock {\em arXiv preprint arXiv:2303.00332}, 2023.

\bibitem{wang2019neuralsourcefilterbasedwaveformmodel}
Xin Wang, Shinji Takaki, and Junichi Yamagishi,
\newblock ``Neural source-filter-based waveform model for statistical parametric speech synthesis,'' 2019.

\bibitem{zhang2024gtsinger}
Yu~Zhang, Changhao Pan, Wenxiang Guo, et~al.,
\newblock ``Gtsinger: A global multi-technique singing corpus with realistic music scores for all singing tasks,''
\newblock {\em Advances in Neural Information Processing Systems}, vol. 37, pp. 1117--1140, 2024.

\bibitem{zhang2022m4singer}
Lichao Zhang, Ruiqi Li, Shoutong Wang, et~al.,
\newblock ``M4singer: A multi-style, multi-singer and musical score provided mandarin singing corpus,''
\newblock {\em Advances in Neural Information Processing Systems}, vol. 35, pp. 6914--6926, 2022.

\bibitem{wang2022opencpop}
Yu~Wang, Xinsheng Wang, Pengcheng Zhu, et~al.,
\newblock ``Opencpop: A high-quality open source chinese popular song corpus for singing voice synthesis,''
\newblock {\em arXiv preprint arXiv:2201.07429}, 2022.

\bibitem{huang2021multi}
Rongjie Huang, Feiyang Chen, Yi~Ren, et~al.,
\newblock ``Multi-singer: Fast multi-singer singing voice vocoder with a large-scale corpus,''
\newblock in {\em ACM MM}, 2021, pp. 3945--3954.

\bibitem{liu2022learning}
Jinglin Liu, Chengxi Li, Yi~Ren, Zhiying Zhu, and Zhou Zhao,
\newblock ``Learning the beauty in songs: Neural singing voice beautifier,''
\newblock {\em arXiv preprint arXiv:2202.13277}, 2022.

\bibitem{liu2022diffsinger}
Jinglin Liu, Chengxi Li, Yi~Ren, Feiyang Chen, and Zhou Zhao,
\newblock ``Diffsinger: Singing voice synthesis via shallow diffusion mechanism,''
\newblock in {\em AAAI}, 2022, vol.~36, pp. 11020--11028.

\bibitem{wang2023mel}
Ju-Chiang Wang, Wei-Tsung Lu, and Minz Won,
\newblock ``Mel-band roformer for music source separation,''
\newblock {\em arXiv preprint arXiv:2310.01809}, 2023.

\bibitem{lee2023bigvgan}
Sang gil Lee, Wei Ping, Boris Ginsburg, Bryan Catanzaro, and Sungroh Yoon,
\newblock ``Big{VGAN}: A universal neural vocoder with large-scale training,''
\newblock in {\em The Eleventh International Conference on Learning Representations}, 2023.

\end{thebibliography}

\end{document}